\documentclass[rmp,twocolumn]{revtex4}
\usepackage{amssymb,amsfonts,amsmath,bm,graphics,epsfig}
\begin{document}

\title{Quantum mechanics as a consistency condition on initial and final boundary conditions}

\author{D. J. Miller}

\affiliation{Centre for Time, Department of Philosophy, University of Sydney NSW 2006, Australia and School of Physics, University of New South Wales, Sydney NSW 2052, Australia}
\begin{abstract}
If the block universe view is correct, the future and the past have similar status and one would expect physical theories to involve final as well as initial boundary conditions. A plausible consistency condition between the initial and final boundary conditions in non-relativistic quantum mechanics leads to the idea that the properties of macroscopic quantum systems, relevantly measuring instruments, are uniquely determined by the boundary conditions. An important element in reaching that conclusion is that preparations and measurements belong in a special class because they involve many subsystems, at least some of which do not form superpositions of their physical properties before the boundary conditions are imposed. It is suggested that the primary role of the formalism of standard quantum mechanics is to provide the consistency condition on the boundary conditions rather than the properties of quantum systems. Expressions are proposed for assigning a set of (unmeasured) physical properties to a quantum system at all times. The physical properties avoid the logical inconsistencies implied by the no-go theorems because they are assigned differently from standard quantum mechanics. Since measurement outcomes are determined by the boundary conditions, they help determine, rather than are determined by, the physical properties of quantum systems.\\[0.3in]
\textit{E-mail address:} D.Miller@unsw.edu.au
\end{abstract}
\maketitle
\noindent
{\bf 1. Introduction}\\

There are two broad approaches to a theory of time. In the A-theory or ``dynamic" view, the past and future have different status and usually the present is thought of as the temporary interface where the future is changing into the past. In the B-theory or ``static" or block universe view, the past and future have similar status and the present is not necessarily a privileged element in the structure of spacetime. There is a view that the special and general theories of relativity are more consistent with the static rather than the dynamic view of time and perhaps even require the static view of time (Yourgrau, 1991).

On the usual account of classical physics, present properties depend on an initial boundary condition (IBC) and time evolution from the IBC or Cauchy data\footnote{In the following, an initial boundary condition means the specification of the Cauchy data required for the solution of the relevant differential equations. In classical mechanics, the differential equations are second order in time and the Cauchy data is the positions and momenta (or two other generalised coordinates) specified over some surface in phase space at some time. The Schr\"{o}dinger equation of quantum mechanics is first order in time and the Cauchy data is the wave function (or more generally, a density operator in the Hilbert space) which specifies at most a precise value for one generalised coordinate.} to the present. This ``single-boundary" formulation seems suitable for physics if the dynamic view of time is correct. The classical problem can also be dealt with as a boundary value problem by dividing the Cauchy data and specifying the parts over each of two surfaces in phase space at two different times. If one of those times is in the future, this ``double-boundary" formulation seems suitable for physics only if the static view of time is correct. In classical physics, the two formulations are equivalent except in certain cases (Costa, Domenech \& Yastremiz, 1999). It has been be argued that the variational principle in classical mechanics is more consistent with the double-boundary formulation but the argument is not decisive (Yourgrau \& Mandelstam, 1968).

In classical physics all the properties of the physical system are determined uniquely by the Cauchy data on one boundary condition (BC) and the time evolution of the system. Therefore, inconsistent physical properties would result from a double-boundary formulation in which the full Cauchy data were specified at two times (unless the data was equivalent). In quantum mechanics, not all the properties of the physical system are determined by the Cauchy data and the time evolution of the system. Therefore, as discussed further in Section 2.1, it is possible to specify full Cauchy data, i.e. the quantum state, independently at the two BC's without necessarily producing any inconsistency. Consequently, unlike the classical case, a double-boundary formulation of quantum mechanics can be a different theory in the sense of leading to the prediction of different properties. For example, at this Conference Sutherland (2005) proposed a double-boundary formulation, in the present sense, of the de Broglie-Bohm theory which leads to quite different trajectories from the conventional single-boundary formulation of de Broglie-Bohm theory.   

It is possible that double-boundary formulations are not just an option for correctly formulating quantum mechanics in our universe but the only correct way of doing so. If the only correct formulation of quantum mechanics involves the double-boundary formulation, one would expect to encounter phenomena that are inexplicable on the basis of a single-boundary formulation. There are no such events in classical physics but there are in quantum physics. For example, the Bell (1987, pp. 14-21) inequalities demonstrate that the correlations in properties of entangled quantum systems are too strong to be explained by the properties of each quantum system conditionalised on its past. This can be recognised as experimental evidence that conditionalising on an IBC is inadequate and that the observed correlations are the result partly of a final boundary condition (FBC). If that is the case, a double-boundary formulation of quantum mechanics is the only correct formulation.

There are other phenomena for which the single-boundary formulation of standard quantum mechanics (SQM)\footnote{The term SQM is meant to refer to the theory presented by von Neumann (1955) or standard text-books on quantum mechanics.} does not provide a physical explanation. For example, there is no consensus on why a quantum system and a measuring instrument do not remain in an entangled superposition of the states corresponding to the possible measurement outcomes after a measurement. If one is willing to accept a many-worlds or relative state point of view, decoherence theory (Zurek, 2003) provides an explanation within the unitary time evolution theory of SQM (process {\bf 2.} of von Neumann (1955, pp. 347-358) as to why the possible measurement outcomes define a preferred basis and why an observer may have a subjective experience of a single outcome. If one seeks a single-world explanation, it is necessary to retain what von Neumann (1955, pp. 347-358) called process {\bf 1.} to select {\it one} of the measurement outcomes as actual and therefore the new state of the quantum system (Busch et al, 1991; Omn\`{e}s, 1999). One motivation for the present work is to show that the objectification of reality can follow from a double-boundary formulation of quantum mechanics without process {\bf 1.} because, for the relevant cases as discussed in Section 2.2, the FBC picks out one only of the measurement outcomes. 

A second motivation for the present work is that SQM has become a theory for predicting the
outcomes of
measurements of quantum systems and it is often said that it is meaningless to ask if quantum
mechanics corresponds to reality in any other respect (Hawking, 1997). There are other versions of quantum mechanics, for example some modal
interpretations, which aim to
deal ``with what there \textit{is}, even in situations in which no
measurements are made'' (Dieks, 1994, emphasis in the original). The no-go theorems (Mermin, 1993) show that
logical contradictions can result if unmeasured properties
are assigned values in accordance with the algorithm of SQM. It follows that either one chooses to remain
silent about the unmeasured properties of a quantum system
(Peres, 1978), or one must develop an algorithm for the assignment of unmeasured properties which is different from the algorithm of SQM for the assignment of measured properties. In a theory based on the dynamic view, the latter option is precluded because properties cannot be assigned in anticipation of whether or not the quantum system is going to be measured. On the other hand, in a theory based on ``advanced action", there is the general possibility that the measurement itself helps determine properties prior to the measurement in the same way that the preparation in SQM determines properties which follow the preparation. Consequently measured and unmeasured properties are not assigned in the same way. For that reason, the no-go theorems can be avoided. The statistical algorithm for assigning values to unmeasured properties without logical inconsistencies in the present time-neutral theory of quantum mechanics is provided in Section 3.2. 

In summary, there are two major topics of the present work. The first is a re-consideration of SQM with an IBC, as usual, plus a non-equivalent FBC as well. This requires the consideration of consistency conditions in Section 2.1 and the consequences are discussed in Section 2.2. The most significant part of this re-formulation is that some phenomena, including measurement outcomes, depend directly on the FBC. The second topic is the definition in Section 3.1 of a set of properties for a quantum system which can be assigned at all times by the algorithm described in Section 3.2. The first topic is independent of the second but the second would not be possible without the first. The relationship of the present theory to previous published work is postponed to Section 4.\\

\noindent
{\bf 2. Initial and final boundary conditions}\\

\noindent
\textit{2.1 Consistency conditions}\\

This work falls into the general category of ``advanced action" approaches in quantum mechanics (Price, 1996) and into the more specific category in which the properties of a quantum system depend, in one form or another, on a state evolved forwards from an IBC and a state evolved backwards from an FBC. Of course, the terms ``initial", ``final", ``forwards" and ``backwards" are used here for convenience only, without any implication about a preferred ``direction of time". As already mentioned, nothing is gained from an advanced action theory of this type if the FBC at time $t_f$ is simply the IBC at time $t_i$ evolved forwards to $t_f$ (or vice versa). The interesting case is when the IBC and the FBC specify different information.
 
The specification of BC's may be constrained by consistency conditions in different ways. Firstly the BC's must be consistent with the physical theory. For example, if the Cauchy data for a problem in mechanics specified the positions and velocities of a set of particles, only velocities less than the speed of light $c$ would be consistent with special relativity. More importantly in the present context is the possibility of overdetermination of the BC's. If the same information required by the Cauchy data is merely {\it distributed} between the IBC's and the FBC's there is no additional constraint. If a full set of Cauchy data is to be specified at both the IBC and the FBC, a consistency condition is vital.

In quantum mechanics, the Cauchy data at some $t$ is the wave function in the case of the Schr\"{o}dinger formulation or, more generally, a density operator in some Hilbert space appropriate to the physical problem of interest. Once specified at $t$, the density operator, i.e. the relevant Cauchy data, is unambiguously determined by unitary time-evolution at any other time before or after $t$ just as in classical mechanics. It might appear that once again we have no discretion in choosing an FBC which is not equivalent to the IBC. That is not true because SQM allows for a change in the density operator which is distinct from unitary time evolution without producing any inconsistencies. The obvious example is von Neumann's process {\bf 1.}, referred to earlier. Therefore, in SQM as opposed to classical mechanics, we can expect to be able to specify an FBC which contains different information from the IBC.

Let us say the IBC for the quantum system is specified by the density operator $\hat{\rho}_i(t_i)$ at time $t_i$ and the FBC is specified by the density operator $\hat{\rho}_f(t_f)$ at time $t_f$. We have that
\begin{eqnarray}
\hat{\rho}_i(t_f) = \hat{U}(t_f,t_i) \hat{\rho}_i(t_i)\hat{U}(t_i,t_f) \mbox{  and  } \nonumber \\ 
\hat{\rho}_f(t_i) = \hat{U}(t_i,t_f) \hat{\rho}_f(t_f) \hat{U}(t_f,t_i)
\end{eqnarray}
where $\hat{U}(t_f,t_i) = \hat{U}^{\dagger}(t_i,t_f) $ is the time-evolution operator determined by the Hamiltonian for the system between the specified times. The density operators are a weighted sums of unique sets of projection operators $\hat{P}_{\alpha} (t_i)$ and $\hat{P}_{\beta} (t_f)$ which project onto subspaces of the Hilbert space:
\begin{equation}
\hat{\rho}_i(t_i) = \sum_{\alpha} a_{\alpha} \hat{P}_{\alpha} (t_i)  \mbox{   and   }
\hat{\rho}_f(t_f) = \sum_{\beta} b_{\beta} \hat{P}_{\beta} (t_f)
\end{equation}
where the $a_{\alpha}$ are real positive numbers which sum to unity and similarly for the $b_{\beta}$. It seems physically reasonable that $\hat{\rho}_f(t_f)$ could only involve those parts of Hilbert space which are reached by the time evolution of $\hat{\rho}_i(t_i)$, that is that the set $\{\hat{P}_{\beta} (t_f)\}$ project onto a union of subpsaces of the set $\{\hat{P}_{\alpha} (t_f)\}$.\footnote{The relationship between $\hat{\rho}_i(t_i)$ and $\hat{\rho}_f(t_f)$ has been discussed in the context of the histories formalisms (Griffiths, 1984; Gell-Mann \& Hartle, 1994; Craig, 1996) without invoking the requirements imposed in this section; see also Section 4 below.} This point is illustrated in Fig.~1. Therefore we are led to a consistency condition on the BC's which is similar to the L\"{u}ders rule for a measurement
\begin{equation}
\hat{\rho}_f(t_f) = B^{-1} \hat{P}_b(t_f) \hat{\rho}_i(t_f) \hat{P}_b(t_f)
\end{equation}
where $\hat{P}_b(t_f)$ is a projection operator such that $B = \mbox{Tr}( \hat{\rho}_i(t_f) \hat{P}_b(t_f)) \neq 0$. The two appearances of $\hat{P}_b(t_f)$ and the factor $B^{-1}$ ensure the right-hand side of Eq.~(3) is a density operator.

The block universe view suggests that the two BC's should be treated on the same footing so we require an expression of the same form for  $\hat{\rho}_i(t_i)$: 
\begin{equation}
\hat{\rho}_i(t_i) = A^{-1} \hat{P}_a(t_i)\hat{\rho}_f(t_i) \hat{P}_a(t_i) 
\end{equation}
where $\hat{P}_a(t_i)$ is a projection operator such that $A = \mbox{Tr}( \hat{\rho}_f(t_i) \hat{P}_a(t_i)) \neq 0$.

It is also necessary that if the initial state is evolved forwards and projected as in Eq.~(3) to give the final state and that state is then evolved back and projected as in Eq.~(4), we must get the same initial state back again. The final requirement is then\footnote{In Eqs. (5) and (6), all operators in the equations for $\hat{\rho}_i$ ($\hat{\rho}_f$) are evolved to $t_i$ ($t_f$).} 
\begin{equation}
\hat{\rho}_i =  (AB)^{-1} \hat{P}_a \hat{P}_b \hat{\rho}_i \hat{P}_b \hat{P}_a
\mbox{  and  }
\hat{\rho}_f = (AB)^{-1} \hat{P}_b \hat{P}_a \hat{\rho}_f \hat{P}_a \hat{P}_b. 
\end{equation}
Since the projection operators are idempotent, it follows from Eq.~(4) or Eq.~(5) that
\begin{equation}
\hat{\rho}_i = \hat{P}_a \hat{\rho}_i \hat{P}_a
\mbox{  and  }
\hat{\rho}_f =  \hat{P}_b \hat{\rho}_f \hat{P}_b . 
\end{equation}
Eqs. (4) - (6) require either that $\hat{\rho}_i = \hat{\rho}_f$ and $\hat{P}_a = \hat{P}_b = \hat{I}$ where $\hat{I}$ is the identity operator, or that
\begin{equation}
\hat{\rho}_i = \hat{P}_a = |a\rangle \langle a| \mbox{  and  } \hat{\rho}_f = \hat{P}_b = |b \rangle \langle b| ,
\end{equation}
i.e. the initial and final states are pure states in the form of normalised rays (one-dimensional (1D) projection operators) in the Hilbert space. The former alternative corresponds to SQM (without process {\bf 1.}) so we will adopt the latter alternative, given by Eq.~(7). Note that the normalisation conditions for the two equations in Eq.~(5) require that $AB = \mbox{Tr}( \hat{\rho}_i \hat{P}_b) \mbox{Tr}( \hat{\rho}_f \hat{P}_a) = {\mbox{Tr}( \hat{P}_b \hat{\rho}_i \hat{P}_b \hat{P}_a)} = {\mbox{Tr}( \hat{P}_a \hat{\rho}_f \hat{P}_a \hat{P}_b)}$ which is satisfied with $\hat{\rho}_i$ and $\hat{\rho}_f$ given by Eq.~(7). 
 
We need to consider quantum systems which can be thought of as an assembly of subsystems. In that case the appropriate Hilbert space is a tensor product space of the Hilbert spaces of the subsystems. In the tensor product space, there are two types of pure states in the form of rays in the space: ones which can be written as a single tensor product of rays in the Hilbert spaces of the subsystems and ones which cannot (Hughes, 1989, pp. 148-151). The latter are the entangled states. We make the assumption that $\hat{\rho}_i$ and $\hat{\rho}_f$ are pure tensor product states, so
\begin{equation}
\hat{\rho}_i = \hat{P}_i^1 \otimes \hat{P}_i^2 \otimes \ldots \hat{P}_i^n \mbox{  and  } \hat{\rho}_f = \hat{P}_f^1 \otimes \hat{P}_f^2 \otimes \ldots \hat{P}_f^n 
\end{equation}
where $\hat{P}_i^j$ and $\hat{P}_f^j$ are the initial and final states (1D projection operators) of subsystem $j$.\\  

\noindent
\textit{2.2 Boundary conditions and measurement}\\
 
Having introduced the concept of IBC and FBC, it is appropriate to ask whether there are any observable consequences of the concept. To show that there are consequences, it is sufficient to consider initially a quantum system in a 2D Hilbert space. Three cases need to be distinguished. They depend on whether the states specified in those components of the IBC and FBC involving the quantum system are put into superpositions or not by the action of the Hamiltonian operating on the system and, if so, how many superpositions occur. For the purposes of illustration, we can visualise the superpositions being due to passage through beam splitters, as illustrated in Fig.~2. The sequence of times are shown along the horizontal axis, which can be read either from the left or the right since no direction of time is preferred. In Fig.~2, the forward (full-lines) evolution of $\hat{\rho_i}(t_i)$ and the backward (dotted-lines) evolution of $\hat{\rho_f}(t_f)$ are shown in the interval between $t_i$ and $t_f$. The initial state (IBC) $\hat{\rho_i} = \hat{P_a}$ is shown on the extreme left as a result of projection by $\hat{P_a}$, indicated by the short full line, of the final state evolved back to the initial time $t_i$. The final state (FBC) $\hat{\rho_f} = \hat{P_b}$ on the extreme right is the result of projection by $\hat{P_b}$, indicated by the short dashed lines, of the initial state evolved to the final time $t_f$.\\  

\noindent
\textit{2.2.1 No superposition of states}\newline
\hspace*{3mm}The trivial case of free evolution is shown in Fig.~2(a). In this case, the state of the quantum system is unique throughout and the projections at $t_i$ and $t_f$ add nothing to SQM.\\

\noindent
\textit{2.2.2 One superposition of states}\newline
\hspace*{3mm}The first interesting case is shown in Fig.~2(b) where there is an interaction at $t_s$ between the quantum system and a beam-splitter (BS). The interaction can be represented by the evolution $| a \rangle \rightarrow \cos \alpha |d \rangle + i \sin \alpha | c \rangle$. The FBC is shown as selecting one of the two paths open to the quantum system after the BS\footnote{Why the projection is onto one of these two paths raises the preferred basis problem which will be discussed in Section 3.1.} in accordance with Eq.~(3) with $\hat{P}_b = |c \rangle \langle c |$. The time reversal of the interaction with the BS leads to two possible paths for the backward evolution of the final state, one of which is the path taken by the initial state because the interaction at the BS is unitary. The projection at $t_i$ in accordance with Eq.~(4) with $\hat{P}_a = |a \rangle \langle a|$ selects the initial state, leading to a consistent picture.

Obviously, the quantum systems which make up the BS, or whatever else causes the superposition are not shown in Fig.~2 but they too, of course, must be subject to IBC's and FBC's.  We have made the usual assumption that the BS is massive enough to be effectively in the same state whichever of the two paths is taken by the quantum system. Therefore, for the purposes of this illustration, the evolution of the BS corresponds to the trivial case shown in Fig.~2(a). \\

\noindent
\textit{2.2.3 Two (or more) sequential superposition of states}\newline
\hspace*{3mm}Fig.~2(c) shows two superpositions in sequence such that the BS's constitute a Mach-Zehnder interferometer (MZI). The significant new element is that the BC's do not determine uniquely the intermediate states between the two BS's. The general conclusion is that the IBC and FBC, in the form presently proposed, uniquely determine the properties of the quantum system (in the preferred basis to be discussed) in the interval between the BC and the {\it nearest interaction in time} which leads to a superposition of states. Although this seems to be a weak conclusion, the next example shows that it has significant consequences.\\

\noindent
\textit{2.2.4 Two sequential superposition of states plus measurement}\newline   
\hspace*{3mm}Figs.~2(e) and (f) are the same as Fig.~2(c) except that a second quantum system $M$ interacts with the first while it is in the MZI. The evolution of $M$ alone is shown in Fig.~2(d). This is a model on the present picture of a measurement at $t_m$ of the first quantum system by a measuring instrument $M$. Here and in the following, the term ``measuring instrument" includes the environment. Before $t_m$, the composite state of the two quantum systems is the pure state
\begin{equation}
| \psi \rangle = |a \rangle  \otimes |M_o \rangle  = (\cos \alpha |d \rangle + i \sin \alpha | c \rangle) \otimes |M_o \rangle .
\end{equation}
As usual, the interaction on path $d$ is assumed to result in an entangled superposition of states 
\begin{equation}
| \psi \rangle = \cos \alpha |d \rangle  \otimes |M_d \rangle  + i \sin \alpha |c \rangle  \otimes |M_o \rangle 
\end{equation}
where the nature of the interaction is such that the state of $M$ changes to $M_d$ if the quantum system is on path $d$ and remains unchanged if it is on path $c$.
  
There are two possibilities which depend crucially on whether or not there is a superposition of the subsequent evolutions of the states states $|M_o \rangle$ and $|M_d \rangle$. In Fig.~2, for illustrative purposes, $M$ is assumed to be composed of two subsystems with states $|m_o^1 \rangle$ and $|m_o^2 \rangle$ for the state $|M_o \rangle $ and $|m_d^1 \rangle$ and $|m_d^2 \rangle$ for $|M_d \rangle $. The question then becomes whether the component subsystems of $M$ form superpositions or not. In Fig.~2(d), $|m_o^1 \rangle$ is shown as not forming superpositions with $|m_d^1 \rangle$ before $t_f$ but $|m_d^2 \rangle $ and $|m_d^2 \rangle$ do form superpositions. Consequently, while the final projection of subsystems 2 does not distinguish between $M_o$ and $M_d$, the final projection of subsystem 1 does do so, as can be seen from Fig.~2(d).

For a real case, the number of subsystems would be at least twenty-odd orders of magnitude larger than in Fig.~2 and the evolution between $t_m$ and $t_f$ would be unimaginably complicated. However the consequence of an argument like the one in the preceding paragraphs remains the same. We can divide the complicated array of final states into what we will call the descendants of $|M_o \rangle $ and $|M_d \rangle $. The descendants of a state are the set of states, including the states of its subsystems, that it evolves into. The descendants must account for the energy, momentum, angular momentum, etc of the original state. By tracking those quantities it is possible in principle to keep account of the descendants of a state even if particle creation and annihilation are allowed for. Therefore it is possible in principle at least to envisage tracking the descendants over cosmological time scales. Of course, the descendants of one measurement outcome are likely to include a huge number of future measurements.

We can now show that the above scheme leads to the same probability for measurement outcomes as SQM. To do so, consider a quantum system and a measuring instrument, originally in the state
\begin{equation}
|M_o \rangle | a \rangle = |M_o \rangle \sum_{i=1}^k \mu_i |q_i \rangle 
\end{equation}
which evolves in the standard manner to $ \sum_{i=1}^k \mu_i |q_i \rangle |M_i \rangle $ as the result of a measurement of observables $\hat{Q}$ with eigenstates $q_i$, some of which may be degenerate. We separate the states of the measuring instrument into those subsystems, represented by $| m_i \rangle $, which are not put in a superposition prior to the FBC and those subsystems, represented by $| n_i\rangle $, which are put into superposition, not necessarily the same for each subsystem. (Also, in this example, assume that the quantum system is put into a superposition.) Then the evolution of the quantum system and measuring instrument is as follows
\begin{eqnarray}
| a \rangle |M_o \rangle  \rightarrow  \sum_{i=1}^k \mu_i |q_i \rangle |M_i \rangle \equiv  \sum_{i=1}^k \mu_i |q_i \rangle |m_i \rangle |n_i \rangle \\
 \rightarrow  \sum_{i=1}^k \mu_i \sum_{\alpha} c_{i \alpha}|q^f_{\alpha} \rangle \sum_{\beta} d_{i \beta} |m^f_{\beta} \rangle \sum_{\gamma} e_{i \gamma}|n^f_{\gamma} \rangle  
\end{eqnarray}
where
\begin{equation}
\sum_{\alpha} |c_{i \alpha}|^2 = \sum_{\beta} |d_{i \beta}|^2 = \sum_{\gamma} |e_{i \gamma}|^2 = 1. 
\end{equation}
Here $i$ labels the descendants and the $|q^f_{\alpha} \rangle $, $|m^f_{\beta} \rangle $ and $|n^f_{\gamma} \rangle $ are the final states of each of the subsystems. In general the final states will be entangled with other quantum systems but that has not been shown in Eq.~(13) to avoid complicating the expressions unduly. Since it has been assumed that the subsystem whose states were $|m_i \rangle $ are not put into a superposition, the final states $|m^f_{\beta} \rangle $ are arranged in sets $\{\beta \}_i$, with no members in common, such that $d_{i \beta}=0 $ unless $\beta \in \{\beta \}_i$. The components of the projection operator $\hat{P}_b$ involved in determining the FBC for $M$ and the quantum system is $\hat{P}_{b'} = | b' \rangle \langle b' |$ where $| b' \rangle $ is one of tensor product states of the component subsystems: $|b' \rangle = |q^f_{\alpha '} \rangle |m^f_{\beta '} \rangle |n^f_{\gamma '} \rangle $. 

Although the FBC is a unique event, the probability for a projection like $\hat{P}_{b'}$ has meaning if the same measurement is performed with as far as possible identically prepared quantum systems and measuring instrument plus environment. Then a projection of the form $\hat{P}_{b'}$ will occur in each repetition and we can seek the probability for the occurrence of $\hat{P}_{b'}$ in the set of repetitions. Since we are seeking a probability measure in a Hilbert space, Gleason's theorem applies (see Hughes, 1989, pp. 146-148) and the probability that $\hat{P}_{b'}$ is the relevant component of $\hat{P}_b$ given the state in Eq.~(13) must be $\mbox{Tr}((|M_o \rangle |a \rangle \langle a| \langle M_o|)| b' \rangle \langle b'|)  = | \mu_j c_{j \alpha '} d_{j \beta '} e_{j \gamma '}|^2$. Because of the normalisation conditions like those given in Eq.~(14) will apply in each repetition, in a large number of runs the probability of outcome $M_j$ will be $|\mu_j|^2$ as predicted by SQM.\\

\noindent
\textit{2.3 What is a measuring instrument?}\\

The important consequence of the previous section is that when, for at least one of subsystems, the descendant states of the possible measurement outcomes do not form a superposition before $t_f$, the outcome of the measurement is uniquely determined by the projection onto the final pure state. It is possible to use that idea to revise the concept of measurement.
 
Let us assume for the time being that a preferred basis set $\{q_i\}$ can be identified and that the IBC projects onto one of the $q_i$ and that a preferred basis set $\{q_f\}$ can be identified and that the FBC projects onto one of the $q_f$. The state of a quantum system at any time $t$ can be expressed in terms of the time evolution to $t$ of the preferred basis $q_i$ or the preferred basis $q_f$. Then the important consequence of the previous section is that quantum systems at time $t$ can be classified into two types: type I at $t$ or type II at $t$. If no superposition of the $q_i$ occurs between the IBC and $t$ and/or if no superposition of the $q_f$ occurs between $t$ and the FBC, the state of the quantum system is uniquely determined by the IBC and/or FBC and the quantum system at $t$, or quantum event, is classified as Type I at time $t$. If a superposition of the $q_i$ does occur between the IBC and $t$ and if a superposition of the $q_f$ occurs between $t$ and the FBC, the state of the quantum system is not uniquely determined by the IBC and FBC and the quantum system is classified as Type II at $t$. If one or more of the subsystems of a composite quantum system are type I at $t$, the composite quantum system is also type I at $t$.

It is clear that the chance that a quantum system is Type I increases with the number of subsystems that compose it. Measuring instruments are prime examples of Type I systems because we know from decoherence theory that the many degrees of freedom of a macroscopic object like a measuring instrument  (including the environment) make it impossible for all practical purposes (FAPP), even deliberately, to form a superposition of the states involved in the possible measuring outcomes because of the loss of phases of the off-diagonal terms in the density matrix. 

With one caveat we can therefore discard measurement as a primitive concept and replace it by the above concepts of Type I and Type II quantum events.  The answer to the question What is a measuring instrument? is that it is a quantum system which is Type I at the time of measurement.

As we have seen, the chances that the FBC will uniquely determine a state increases with the complexity, i.e. the number of subsystems, of the quantum systems but Type I quantum systems are not confined to complex quantum systems. Therefore the answer to the above question will include quantum systems not usually thought of as measuring instruments because, by chance, the conditions are satisfied even when the complexity is small. On the other hand, there will be complex quantum systems for which by chance, the conditions are not satisfied. For the current approach to be acceptable, those latter cases must be sufficiently rare to be subsumed into the category of experimental error.

The caveat mentioned above is that we have so far relied explicitly or implicitly on measurement outcomes to identify the preferred basis used for the projection at the FBC. To eliminate the concept of measurement entirely, there needs to be an independent way of identifying a preferred basis and that is considered in the next section.\\ 

\noindent
{\bf 3. Probabilities for physical properties}\\

In this section we turn to the possibility of assigning unmeasured properties to a quantum system between the preparation and the next measurement of the quantum system. The set of unmeasured properties that can be assigned will be defined in terms of a preferred basis in the next section and then a probability measure for the unmeasured properties will be proposed.\\ 

\noindent
\textit{3.1 Physical basis}\\

The argument in Section 2.2 relied on the projection involved at the FBC (or IBC) distinguishing between the various measurement outcomes, therefore we are assuming there is a preferred basis determined by the states corresponding to the measurement outcomes. At one level, the reason is that decoherence theory shows that the states corresponding to what we recognise as the measurement outcomes form a preferred basis, at least FAPP. That reason suits the present theory for the purposes of what we recognise as measurements.

However the reason is unsatisfactory because it is true only FAPP or if it is asserted in a stronger way, measurement needs to be regarded as a primitive concept. Therefore we propose below an alternative answer. The main motivation for doing so is to pursue the second aim of this work which is to investigate the unmeasured properties of quantum systems. It should be emphasised that the rest of this section is severable from the foregoing sections because we could choose to rely on the preferred basis picked out by decoherence theory for purposes of those sections.

Usually the preferred basis problem is addressed from an abstract point of view, relying on the formal properties of Hilbert space itself and often with the aim of maximising the set of properties that can be assigned. If one starts from a more physical point of view, a prime candidate for a preferred basis are the eigenstates of the complete set of commuting observables (CSCO) which play an important role also in SQM. The eigenstates of a CSCO form an essential role in the ``contextual objectivity" approach to quantum mechanics (Grangier, 2002). 

Often there may not be enough information about the quantum system to identify the CSCO. One way of identifying an observable in a CSCO is by the condition for a quantum system to be ``measurement-ready'' (MR) for an observable (Miller, 2006). Conditions analogous to being MR occur elsewhere in the literature. For example the ``ideal-negative-result'' method of measurement used in the experiments proposed by Leggett (1999), which were also reviewed at this conference (Leggett, 2005), is a similar concept at the macroscopic level. The property of being MR proposed here is also similar to the concept of ``partial measurement'' that has been used in the analysis of weak measurements (Kastner, 2004). 

The idea of being MR is very simple and will be illustrated by reference to the MZI in Fig.~2(c), although the concept is a general one not confined to the example. The quantum system is prepared in the state $|a \rangle $ and is subjected to a physical interaction at time $t_{1}$ (the BS in Fig.~2(c)). The state of the quantum system in a formal sense can be written as $| a \rangle = \cos \alpha | d \rangle + i \sin \alpha | c \rangle $ (see Eq.~(9)) both before or after $t_{1}$ but it seems obvious that it is  physically meaningful to do so after $t_{1}$ but not before. The difference is that after $t_{1}$, but not before, a measurement can be performed which ascertains, at least sometimes, the state of the quantum system {\it without further physical interaction with the quantum system}; an example is Fig.~2(e) where the quantum system is found to be on path $c$ because the measuring instrument does not register. Therefore, while there is nothing special about the $| c \rangle $/$| d \rangle $ basis after $t_{1}$ in a formal sense, the above remarks make it physically reasonable to recognise the $| c \rangle $/$| d \rangle $ basis as a preferred basis.\footnote{The states $|c \rangle $ and $|d \rangle $ would not be regarded as a preferred basis in the consistent histories formalism (Griffiths, 2002, pp. 178-183).} It is easy to show that if a quantum system is in a MR condition for observable $C$ it cannot be in a MR condition for any other observable which does not commute with $C$ (Miller, 2006).

After a measurement, the measurement basis (the basis corresponding to the measurement outcomes, e.g. picked out by decoherence theory) will correspond to the physical basis defined above for the observable that has been measured. That is obvious because in a measurement, the quantum system becomes entangled with the measuring instrument and the state of the quantum system can be determined {\it without further interaction with the quantum system} merely by observing the instrument or performing another measurement on the measuring instrument.

In summary, the present proposal is that any quantum system is in a measurement-ready condition for each of the observables in a CSCO, the CSCO having been picked out by the physical circumstances experienced by the quantum system. The eigenstates of the CSCO (which are non-degenerate) constitute a preferred basis, which will be referred to as the \textit{physical basis}. Finally we assume that the projections $\hat{P}_a$ and $\hat{P}_b$ of Section 2.1 which determine the IBC and the FBC are projections onto the physical basis determined by the CSCO pertaining at the initial and final times respectively.\\ 

\noindent
\textit{3.2 Physical properties}\\

In the previous section, it was shown that for the case of a MR observable, and only in that case, in some repetitions of an experiment to measure that observable, the quantum system can be said to be in an eigenstate of the observable and to possess the corresponding property despite the fact there was no direct interaction with the eigenstate during the measurement. Therefore it is tempting to assert that whenever a quantum system is in a MR condition for an observable, the quantum system is in one of the eigenstates of that observable and, even when no measurement is actually carried out, possesses the corresponding property (which will be called a \textit{physical property} in the following).  

It is not possible to make the latter assertion in SQM because, if as assumed, no measurement is actually carried out, the MR condition is reversible, for example by a choosing the second BS interaction at $t_2$ in Fig.~2(c) as the inverse of the interaction at the first BS. According to SQM, the original state is then restored and that is due to interference among {\it all} the states ($| c \rangle $ and $| d \rangle $ in the example). Given that the next state appears to be the result of interference among all the states, it it seems illogical to assign only one of $| c \rangle $ or $| d \rangle $ to the quantum system between $t_{1}$ and $t_{2}$. But it does not follow in the present theory that the original state is restored in the above circumstances. It is true that the original state is restored in the circumstances that it is measured but that, according to the present approach, is because a measurement outcome is the result of the FBC, not the result of interference among the unpossessed properties of the quantum system. 

The next step is to assign probabilities that a quantum system will posses a physical property or sequence of physical properties. On the present theory, the BC's determine the preparation and measurement outcomes and so for an individual quantum system, a preparation and next measurement represent the effective BC's. Although the IBC and FBC are pure states, and therefore the effective BC's must also be pure states, there may not be enough information to identify them and so the preparation and next measurement should in general be represented by density operators, $\hat{\rho}_p$ and $\hat{\rho}_m$ respectively. The aim then is to find the probability of a sequence of physical properties between a preparation and the next measurement in terms of $\hat{\rho}_p$ and $\hat{\rho}_m$. The following extends earlier work (Miller, 1997, 1998) along those lines.

The possible physical properties are the eigenstates of the observables of the current CSCO. The CSCO will change every time the quantum system encounters a Hamiltonian which does not commute with current set. It is sufficient to develop the theory for one of the observables from the CSCO because each observable in the CSCO is drawn from a set of non-commuting observables which evolve independently of one another. As for the histories formalisms (Griffiths, 2002), the $j$th physical property will be represented by a projection operator from the set $\{\hat{P}_{\alpha_{j}}^{j}\}$ where $\hat{P}_{\alpha_{j}}^{j} \equiv \hat{P}_{\alpha_{j}}^{j}(t_j)$ projects onto the eigenspace, labelled by $\alpha_{j}$, of the observable from the CSCO current at $t_j$. Given $\hat{\rho}_p$ and $\hat{\rho}_m$, the probability is required for the ordered sequence of $k$ physical properties $\alpha ' \equiv \alpha_{1}, \alpha_{2}, \ldots, \alpha_{k}$ out of the set $\{\alpha\}$ of all possible ordered sequences of the physical properties. The probability will involve the corresponding sequence of projection operators in the tensor product Hilbert space $\otimes^n \mathcal{H}$: $\hat{S}_{\alpha '} = \hat{P}_{\alpha_{1}}^{1} \otimes \hat{P}_{\alpha_{2}}^{2}  \otimes \ldots \otimes \hat{P}_{\alpha_{k}}^{k}$ and the chain operator (Griffiths, 2002, pp. 137-140) $\hat{K}_{\alpha '} = \hat{P}_{\alpha_{1}}^{1} \hat{U}(t_1,t_2) \hat{P}_{\alpha_{2}}^{2}  \hat{U}(t_2,t_3) \ldots \hat{U}(t_{k-1},t_k) \hat{P}_{\alpha_{k}}^{k}$ in $\mathcal{H}$. We propose the following expression which satisfies the usual requirements for a probability measure
\begin{eqnarray}
\mbox{Prob}(\hat{S}_{\alpha '}|\hat{\rho}_p, \{\alpha\}, \hat{\rho}_m)  =  \mbox{Tr}_{\otimes^n \mathcal{H}}(\rho[\hat{\rho}_p, \{\alpha\}, \hat{\rho}_m] \hat{S}_{\alpha '}) \;\;\;\;\\ 
\mbox{  where  } \rho[\hat{\rho}_p, \{\alpha\}, \hat{\rho}_m]  =  N^{-1} \sum_{\{\alpha\}} |\mbox{Tr}_{\mathcal{H}}(\hat{\rho}_p \hat{K}_{\alpha}\hat{\rho}_m)| \hat{S}_{\alpha} \;\;\;\; \\
\mbox{  and  } N  =  \sum_{\{\alpha\}}|\mbox{Tr}_{\mathcal{H}}(\hat{\rho}_p \hat{K}_{\alpha}\hat{\rho}_m)| \mbox{Tr}_{\otimes^n \mathcal{H}}(\hat{S}_{\alpha}) .\;\;\;\;
\end{eqnarray} 

The first thing to note is that the density operator in Eq.~(16) depends on the set of physical properties represented by $\{\alpha\}$. Normally that would allow an experiment to be set up for superluminal signalling (Peres, 1993) but that does not apply in the present case because the probability measure is expressly confined to {\it unmeasured} properties. 

As an example we apply the above expression to the case given in Fig.~2(c). We assume the change of physical basis occurs as follows $|a \rangle  \rightarrow  \cos \theta |d \rangle + i \sin \theta |c \rangle $, $|c \rangle  \rightarrow  \cos \phi |f \rangle + i \sin \phi |e \rangle $ and $|d \rangle  \rightarrow  i \sin \phi |f \rangle + \cos \phi |e \rangle $ with $0 \leq \theta, \phi \leq \pi/2$. We assumed previously that the IBC picks out the initial state $|a \rangle $ and the FBC picks out the final state $|e \rangle $ (which is possible provided $\theta + \phi \neq \pi/2$), so $\hat{\rho}_p = 
|a \rangle \langle a |$ and $\hat{\rho}_m = |e \rangle \langle e|$. Sequences involving the projector $|b \rangle \langle b |$ or  $|  f \rangle \langle f |$ obviously have zero probability according to the above equations. Therefore there are only two possible sequences of properties $\{c,d\}$ and 
$\hat{K}_c = |a \rangle \langle a |c \rangle \langle c |e \rangle \langle e | = -\sin \theta \sin \phi |a \rangle \langle e |$ and 
$\hat{K}_d = |a \rangle \langle a |d \rangle \langle d |e \rangle \langle e | =  \cos \theta \cos \phi |a \rangle \langle e |$. From Eq.~(15), we find that Prob$(S_c|(|a \rangle \langle a|, \{c,d\}, |e \rangle \langle e|)) = \sin \theta \sin \phi/ \cos (\theta - \phi) $ and Prob$(S_d|(|a \rangle \langle a|, \{c,d\}, |e \rangle \langle e|)) = \cos \theta \cos \phi  / \cos (\theta - \phi) $ (for the case of the above condition $0 \leq \theta, \phi \leq \pi/2$). This compares with the probability according to SQM for a {\it measurement} on path $c$ of $\sin^2 \theta$ and on path $d$ of $\cos^2 \theta$.\\

\noindent
{\bf 4. Relation to other theories}\\

The present work is another contribution to those approaches to quantum mechanics which rely in one way or another on the future as well as the past. Many of those theories were reviewed at the present conference by those who have proposed them. The main aim of this section is to relate the present approach to those theories, unfortunately in a necessarily very brief and rather cursory way. Recently, Sutherland (1998) has considered a way of extending the general approach to quantum field theory.

A theory of quantum measurement involving {\it two-time boundary conditions} has been worked out in detail by Schulman (1997) and was reviewed at this conference (Schulman, 2005). The present approach is similar in that the ``grotesque" superposition of macroscopic measurement outcomes of SQM is assumed to be avoided by boundary conditions. Major differences are that in the present theory the IBC and FBC are both involved in avoiding the grotesque states and it is assumed the IBC and FBC are related by non-unitary projections. In Schulmann's theory, there are no non-unitary projections, the FBC being the unitary time-evolution of the IBC, and it is the special choice of IBC that avoids the grotesque states, with the FBC being involved only to the extent of making the special choice of IBC more plausible. The rest of the theories in this section, including the present one, take the additional step of conditionalising on both the IBC and FBC, with the two BC's not being the unitary time evolution of each other.

The basic element of the {\it transactional interpretation of quantum mechanics} (Cramer, 1986), which was reviewed at this conference by Cramer (2005), is an exchange of retarded waves from an emitter (preparation event) and advanced waves from an absorber usually taken as the next encounter with a macroscopic object (next measurement). The present approach could be said to involve a similar idea in which the emitter and absorber are translated into the IBC and FBC and the formalism does not rely expressly on the wave solutions to the Schr\"{o}dinger equation. The experimental outcome is the result of the establishment of a ``transaction" in one case and by the projection of the IBC to set the FBC in the other.
    
At this conference, Vaidman (2005) reviewed the {\it two-vector formalism} (see also, for example, (Aharanov \& Vaidman, 2001)) which expressly relies on initial and final states to determine intermediate properties. In terms of the two-vector formalism, Gruss (2000)\footnote{The author thanks Lev Vaidman for drawing his attention to the paper by Gruss (2000) during the conference.} and Aharonov and Gruss (2005) have suggested the main idea that has been arrived at independently in the present work, namely that each measurement outcome can be determined by a properly chosen final boundary condition, thereby solving the measurement problem of SQM.  In both cases, the FBC involves a preferred basis. In Aharonov and Gruss (2005), the preferred basis is a ``classical basis due to the effect of decoherence" while in the present case the primitive concept is the physical basis determined by the MR concept in Sect.~3.1. In the most important case of measurements, the two criteria result in the same basis. The expressions for how the IBC and FBC determine intermediate properties, Eqs.~(15)-(17) above, is different from that proposed in the two-vector formalism (Aharonov \& Vaidman, 2002).

The question of an IBC and FBC has been considered in the theories of consistent histories and decoherent histories (Griffiths, 1984; Gell-Mann \& Hartle, 1994; Craig, 1996). In those contexts, the consistency condition which is usually imposed is that the trace of the product of the initial and final states (density operators) must be non-zero. That condition does not restrict the choice of IBC and FBC to the same extent as additional considerations have done in the present case. It certainly does not lead to the conclusion that the initial and final states must be pure states as concluded here. However it is known in decoherence theory that if the IBC and FBC {\it are} pure states, the decoherence functional factorises with the result that only one, or a maximum of two, coarse-grained histories are possible. Since a sequence of measurements produce a set of coarse-grained histories, it follows (amongst other things) that if the initial and final states are pure states, then the outcomes of all measurements are determined by the IBC and FBC. Within the decoherent histories theory that situation is said to be ``bizarre" (Gell-Mann \& Hartle, 1994, p. 333). From the present point of view it is a distinct advantage that at some level of coarse-graining properties are uniquely determined by the BC's so that a superposition of those (macroscopic) states which constitute the measurement problem in SQM never occurs and that the determinism of classical mechanics is restored to quantum mechanics at the macroscopic level.\\  

\noindent
{\bf 5. Discussion and Conclusion}\\

The starting point for this work has been to investigate the consequences of imposing final as well as initial boundary conditions to give a different theory of non-relativistic quantum mechanics. In a deterministic theory like classical mechanics, a BC at one time determines all the physical properties of the system at all times and therefore a second BC at a different time is redundant. In a non-deterministic theory like SQM, it is possible to impose an FBC which is not simply the time evolution of the IBC. It is for that reason that imposing an FBC in quantum mechanics can result in a different theory. Even though there is some discretion in choosing an IBC and FBC which are not simply time evolutions of each other in quantum mechanics, they must be consistent with each other. The question of consistency was considered in Section 2.1 where it was argued that each BC must be a projection of the other BC (evolved to the time of the first). It then follows that consistency requires that the IBC and FBC must be pure states.

The requirement that the IBC and FBC must be pure states has significant consequences. It was further required that the pure states were (i) a tensor product of the states of the subsystems and (ii) in a preferred basis. The preferred basis consists of the eigenstates of the observables making up a CSCO which is determined by the physical circumstances experienced by the quantum system. In particular, if a quantum system is measured, the measured observable is part of the CSCO until the quantum system experiences a Hamiltonian which does not commute with the observable.

The motivation for this choice of a preferred basis is that it is possible, in principle at least, to place a quantum system in a ``measurement-ready condition" for a CSCO but not for any greater number of observables than make up a CSCO. The appealing feature of an observable of a quantum system being in the measurement-ready condition is that the quantum system can sometimes be measured to possess one of the eigenvalues of that observable without any interaction with the corresponding eigenstate (the interaction necessary for the measurement being with the eigenstates that the quantum system is found not to be in). Therefore it is physically reasonable to suggest that the quantum system has the corresponding property merely because it is in measurement-ready condition, even when it is not measured. 

As discussed in Section 2.3, at any time $t$ there may be quantum systems which are what we have called Type I systems at $t$: those whose possible states at time $t$ do not form superpositions in the interval between $t$ and the time when the FBC is imposed (or between $t$ and the time when the IBC is imposed, although that condition has not been explored to any extent in this work).  

The important point is that the BC's determine uniquely one only of the possible states of Type I quantum systems at $t$. Therefore Type I quantum systems at $t$ possess one only of the possible properties they could have according to SQM at time $t$. For a macroscopic quantum system, especially measuring instruments connected to the environment, it is very likely that at least one of its subsystems will be Type I at $t$. Thus the apparently arbitrary divide between measuring systems and other quantum systems, which has remained a major puzzle of SQM, is explained in terms of the difference between quantum systems whose physical states at the time of measurement respectively do not, or do, form superpositions between $t$ and the IBC or the FBC (are Type I or Type II respectively).

In terms of a many-worlds interpretation, what is being claimed here is that the FBC and IBC pick out just one of the many ``worlds" that are possible according to SQM without process {\bf 1.} (the projection postulate). In the world that is picked out, all measurement outcomes, and some additional phenomena, are uniquely determined. Put in a slightly different way, according to SQM without process {\bf 1.}, the final state of the universe is an extremely complex superposition of states. One can ask, What is the final state of the universe according to SQM \textit{with} process {\bf 1}? Most of the possible states in the first case would be eliminated and all of the remaining states in the final state of the second case would have evolved from unique measurement outcomes selected by process {\bf 1}. In the present theory, the final state is the second one but it is independently arrived at from the consistency condition on the BC's and it helps determine the measurement outcomes by a process of backward causation or advanced action. Instead of process {\bf 1.} being (arbitrarily) invoked at every measurement, an equivalent, and independently justified, process is invoked just once at the BC's.  

Since measurement outcomes are determined by the BC's, the quantum system is relieved of the responsibility of determining measurement outcomes directly and we can explore a new theory for assigning unmeasured properties. On a block universe view, they should depend on the IBC and FBC. Since the latter determine the preparation and measurement states of a quantum system, the preparation and measurement states encode the relevant BC information and we should look for a theory of unmeasured properties, or ``physical properties", conditioned on the preparation and measurement states. In Section 3.2 we have suggested a specific expression for assigning a sequence of physical properties (eigenstates of a CSCO) to a quantum system at all times between preparation and measurement. The next question is why is not such an assignment of properties rules out by the no-go theorems (Mermin, 1993)?

The no-go theorems appear to prevent the assignment of unmeasured properties to a quantum system in a physically reasonable way, i.e. without relying on non-local causation. The essential reason why the present theory can avoid those restrictions is that it does not assign physical properties according to the statistical algorithm of SQM. For a theory based on a dynamic theory of time, this would create a conflict with experiment because, at the discretion of the experimenter, any one of the physical properties could be measured and the result must conform with SQM. Thus it has appeared that all properties assigned (for example by a hidden variable theory) to a quantum system whether measured or not, must be in accordance with SQM on penalty of potentially disagreeing with experiment and then the no-go theorems come into play. On the static or block universe theory of time, this restriction does not apply because a property that is actually measured can be included in the description and perhaps taken into account in assigning the unmeasured properties as is done in Eqs.~(15)-(17).

In summary, in a deterministic theory like classical mechanics, it is possible to define consistently only one set of Cauchy data (IBC) for the differential equations of the theory. In a non-deterministic theory like quantum mechanics, it is possible to define consistently different sets of Cauchy data for the differential equations of the theory at two different times (IBC and FBC). By adopting the latter option in a time-neutral formalism consistent with the block universe view, quantum mechanics becomes deterministic on a macroscopic level and it is possible to assign unmeasured properties (local hidden variables) in a way which avoids the no-go theorems. The ``mysteries" of SQM should not be so regarded; the better view is that they are evidence of a final boundary condition.

\begin{acknowledgments}
I wish to thank participants at the conference and earlier visitors to the Centre for Time for helpful comments and stimulating questions, in particular Harvey Brown in relation to boundary conditions.
\end{acknowledgments}

\begin{figure*}
\includegraphics{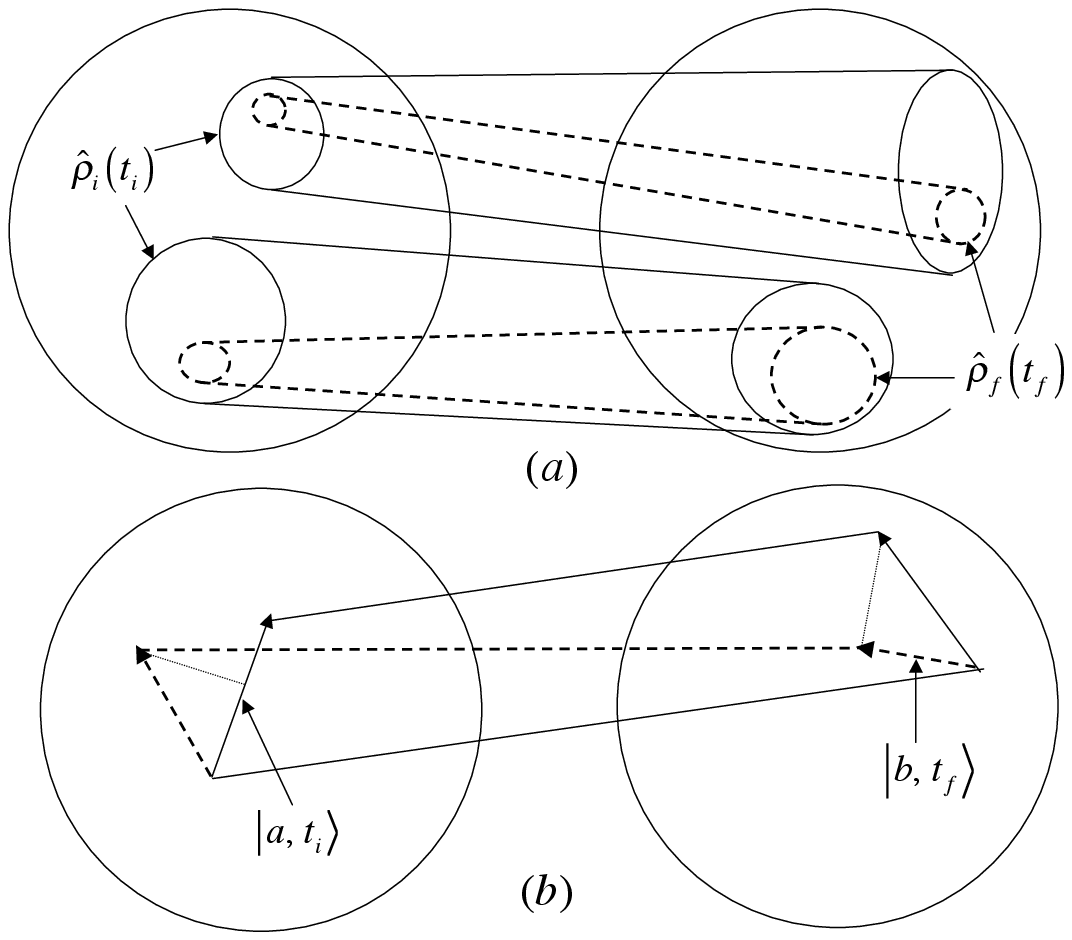}
\caption{(a) The IBC at $t_i$ is shown on the left by the two areas bounded by solid lines which represent schematically a density operator $\hat{\rho}_i(t_i)$  with two eigenstates. The solid lines show the evolution of $\hat{\rho}_i$ to $t_f$ where it is projected to form the FBC in the form of $\hat{\rho}_f(t_f)$ represented by the areas bounded by dashed lines. The evolution of $\hat{\rho}_f$ back to $t_i$ is show by the dashed lines. It is not possible to regain $\hat{\rho}_i$ by a projection of $\hat{\rho}_f$ unless the eigenspaces of $\hat{\rho}_i$ and $\hat{\rho}_f$ have the same dimensionality and if they do, the two must be the same density operator, i.e. the IBC and the FBC are equivalent. (b) The diagram shows that if the IBC and the FBC are pure states, it is possible, subject to re-normalisation, for each to be the projection of the time evolution of the other.}
\label{fig1}
\end{figure*}

\begin{figure*}
\includegraphics{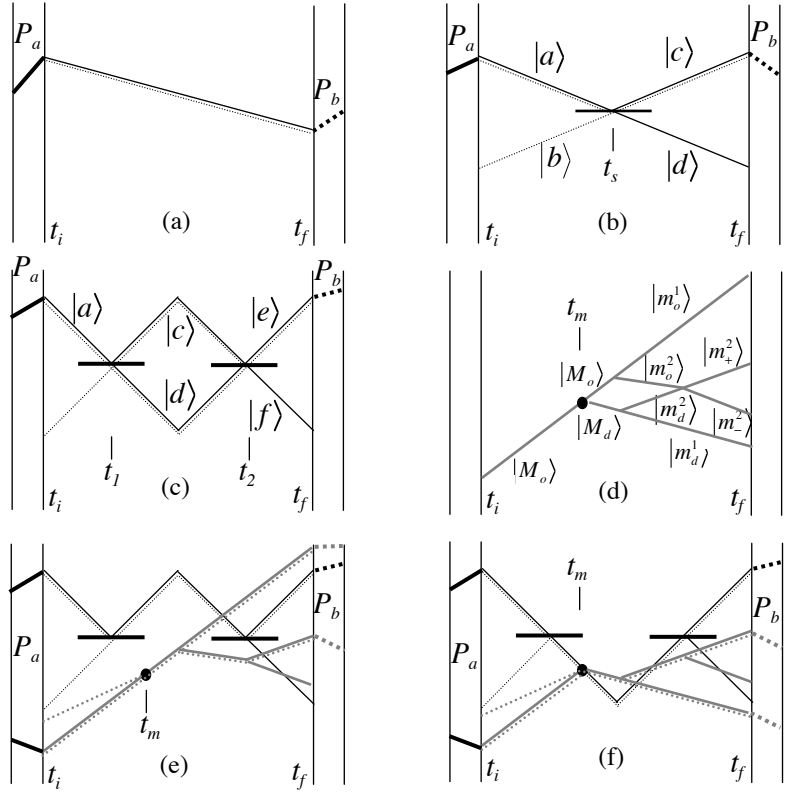}
\caption{Sequences of properties discussed in the text. (a) The trivial case of free evolution. (b) The case of one superposition of states. The IBC and FBC determine a unique path, the one along which the solid and dashed lines coincide. (c) In the case of two superpositions and no measurement, the path between the beam-splitters is not uniquely determined by the IBC and FBC. (d) The sequences of properties for the measuring instrument which is shown measuring the quantum system in (e) and (f). After the measurement shown by the black dot, the state of the measuring instrument remains as $| M_o \rangle $ if the quantum system is on path $|c \rangle $ or changes to $| M_d \rangle $ if the quantum system is on path $|d \rangle $. In either case, the MI is shown as breaking up into two component subsystems, with subsystem 2 subsequently forming a superposition. (e) The case when the FBC projects onto $|m^1_o \rangle $ (and $|m^2_+ \rangle $ and $|e\rangle $).  Because the quantum system and measuring instrument are entangled (see Eq.~(13)), the quantum system must follow path $|c\rangle $. (e) The case when the FBC projects onto $|m^1_d \rangle $ (and $|m^2_+ \rangle $ and $|e\rangle $).}
\label{fig2}
\end{figure*}

\end{document}